\documentclass[francais]{gretsi}

\usepackage[utf8]{inputenc}

\usepackage[T1]{fontenc}

\usepackage{amsmath, amssymb, amsfonts}
\usepackage{indentfirst}
\usepackage{setspace}

\begin{document}


\titre{Super-résolution non supervisée d'images hyperspectrales de télédétection utilisant un entraînement entièrement synthétique}

\auteurs{  
  \auteur{Xinxin}{Xu}{xinxin.xu@telecom-paris.fr}{}
  \auteur{Yann}{Gousseau}{yann.gousseau@telecom-paris.fr}{}
  \auteur{Christophe}{Kervazo}{christophe.kervazo@telecom-paris.fr}{}
  \auteur{Saïd}{Ladjal}{said.ladjal@telecom-paris.fr}{}
}

\affils{  
  \affil{}{LTCI, Télécom Paris, Institut Polytechnique de Paris, 19 Pl. Marguerite Perey, 91120 Palaiseau, France}
}

\resume{La super-résolution mono-image hyperspectrale (SISR) vise à améliorer la résolution spatiale des images tout en conservant leur richesse spectrale. La plupart des méthodes actuelles reposent sur un apprentissage supervisé nécessitant des données de référence haute résolution, souvent indisponibles en pratique. Pour surmonter cette limite, nous proposons une approche d’apprentissage non supervisée basée sur la génération de données synthétiques. L’image hyperspectrale est d’abord décomposée en matériaux et abondances via un algorithme de démélange hyperspectral. Un réseau de neurones est ensuite entraîné à super-résoudre les cartes d’abondance à partir de données synthétiques générées selon un modèle feuilles mortes, qui imitent les propriétés statistiques des abondances réelles. L’image hyperspectrale haute résolution est reconstruite par recombinaison des cartes d’abondance super-résolues et des matériaux. Les résultats expérimentaux confirment l’efficacité de cette méthode et montrent l’intérêt des données synthétiques pour l’entraînement.
}

\abstract{Hyperspectral single image super-resolution (SISR) aims to enhance spatial resolution while preserving the rich spectral information of hyperspectral images. Most existing methods rely on supervised learning with high-resolution ground truth data, which is often unavailable in practice. To overcome this limitation, we propose an unsupervised learning approach based on synthetic abundance data. The hyperspectral image is first decomposed into endmembers and abundance maps through hyperspectral unmixing. A neural network is then trained to super-resolve these maps using data generated with the dead leaves model, which replicates the statistical properties of real abundances. The final super-resolution hyperspectral image is reconstructed by recombining the super-resolved abundance maps with the endmembers. Experimental results demonstrate the effectiveness of our method and the relevance of synthetic data for training.}

\maketitle


\section{Introduction}
Une image hyperspectrale (HSI) est un cube de données contenant des centaines d'images couvrant une large gamme de longueurs d'onde. En raison de leur haute résolution spectrale, les images hyperspectrales sont très utilisées dans de nombreuses applications, telles que le suivi de la végétation, l'utilisation des sols et l'astrophysique \cite{fahes2022unrolling} pour n'en citer que quelques-unes. Toutefois, en raison de contraintes optiques, les capteurs hyperspectraux ont généralement une résolution spatiale plus faible que leurs homologues multispectraux. Par conséquent, la super-résolution des images hyperspectrales, qui vise à augmenter la résolution spatiale des HSIs tout en conservant leur résolution spectrale, a suscité un grand intérêt au cours des dernières décennies. La plupart des méthodes existantes peuvent être classées en deux catégories : la fusion multi/hyperspectrale et la super-résolution d'une image unique (\textit{Single Image Super Resolution} ou SISR).

Bien que les méthodes de fusion multi/hyperspectrales aient été largement étudiées, leur application se heurte à la difficulté de recaler avec précision les paires d'images multi/hyperspectrales. Ainsi, dans cet article, notre attention se porte sur les approches SISR. Initialement, les travaux sur la SISR étaient basés sur des modèles classiques relativement bien compris. Par exemple, en 2005, Akgun et al.\cite{akgun2005super} ont formulé la tâche de super-résolution comme un problème inverse, résolu à l'aide de projections sur des ensembles convexes \cite{bauschke1996projection}. Plus récemment, l'utilisation des réseaux de neurones convolutifs (\textit{Convolutional Neural Networks}, ou CNN) s'est imposée comme une approche très efficace. Par exemple, Yuan et al. \cite{yuan2017hyperspectral} ont appliqué une approche d'apprentissage par transfert, utilisant un réseau préalablement entraîné sur des images naturelles. D'autres travaux ont exploré l'utilisation de convolutions tridimensionnelles, permettant d'extraire à la fois des informations spatiales et spectrales. Cela inclut des convolutions 3D \cite{mei2017hyperspectral}, des combinaisons de convolutions 1D et 2D \cite{li2019dual}, ou un mélange de convolutions 2D et 3D \cite{li2020mixed, wang2020hyperspectral}.

Les méthodes SISR actuelles sont toutefois généralement entravées par le manque de jeux de données d'entraînement supervisé : il n'existe pas, à notre connaissance, de jeux de données avec une vérité terrain de HSI super-résolue, ce qui limite fortement l'utilisation de méthodes supervisées. C'est pourquoi nous proposons une nouvelle méthode pour générer des données synthétiques étiquetées avec les vérités terrain afin d'entraîner les réseaux de super-résolution.

Dans le contexte des images naturelles, de nombreux modèles stochastiques ont été étudiés, tels les champs de Markov \cite{cross1983markov}, les modèles d'ondelettes \cite{heeger1995pyramid}, les modèles gaussiens \cite{galerne2011micro} ou le modèle feuilles mortes \cite{alvarez1999size, cao2010dead, gousseau2003dead}. Un avantage de ce dernier est de permettre simplement, avec peu de paramètres, un bon respect de nombreuses statistiques des images, en particulier non-gaussiennes. Récemment, ces propriétes ont été mises à profit pour la génération de données permettant d'entraîner efficacement des réseau neuronaux de restauration \cite{achddou2023fully}. Dans cette direction, nous avons développé une nouvelle méthode de SISR non supervisée, reposant sur un apprentissage entièrement synthétique basé sur le modèle des feuilles mortes \cite{xu2024unsupervised}. Dans cette communication, nous propoonse une version plus aboutie de notre approche , intégrant une étape de démélange hyperspectral. Les sections suivantes détaillent notre méthode ainsi que les résultats obtenus.

\section{La méthode proposée}

\subsection{Aperçu général}

Pour réaliser la super-résolution, l'objectif est d'obtenir une image hyperspectrale à haute résolution spatiale $HSI_{HR} \in \mathbb{R}^{L \times H \times W}$ à partir d'une image à basse résolution, $HSI_{LR} \in \mathbb{R}^{L \times h \times w}$, où $h$, $w$, $H$ et $W$ ($h < H$ et $w < W$) sont les dimensions spatiales, et $L$ est la dimension spectrale. En dépit de sa simplicité et des limites associées \cite{kervazo2021provably}, nous faisons l'hypothèse que chaque pixel dans la HSI peut être décomposé exactement en utilisant le modèle de démélange linéaire hyperspectral suivant :
\begin{equation}
    HSI_{LR}(l,i,j) = \sum_{n=1}^{N} S(l,n) \cdot A_{LR}(n,i,j)
    \label{eq:unmixing}
\end{equation}
où $A_{LR} \in \mathbb{R}^{N \times h \times w}$ et $S \in \mathbb{R}^{L \times N}$ représentent respectivement les cartes d'abondances et les matériaux de la HSI à basse résolution, et $N$ est le nombre de matériaux. En pratique, nous appliquons la méthode du démélange hyperspectral par Volume Minimum sur $HSI_{LR}$ \cite{gillis2014successive} pour extraire les matériaux $S$. Ensuite, afin de limiter les erreurs de reconstruction, les abondances $A_{LR}$ sont obtenues en utilisant la méthode des moindres carrés.

\begin{figure}
    \centering
    \includegraphics[width=1\linewidth]{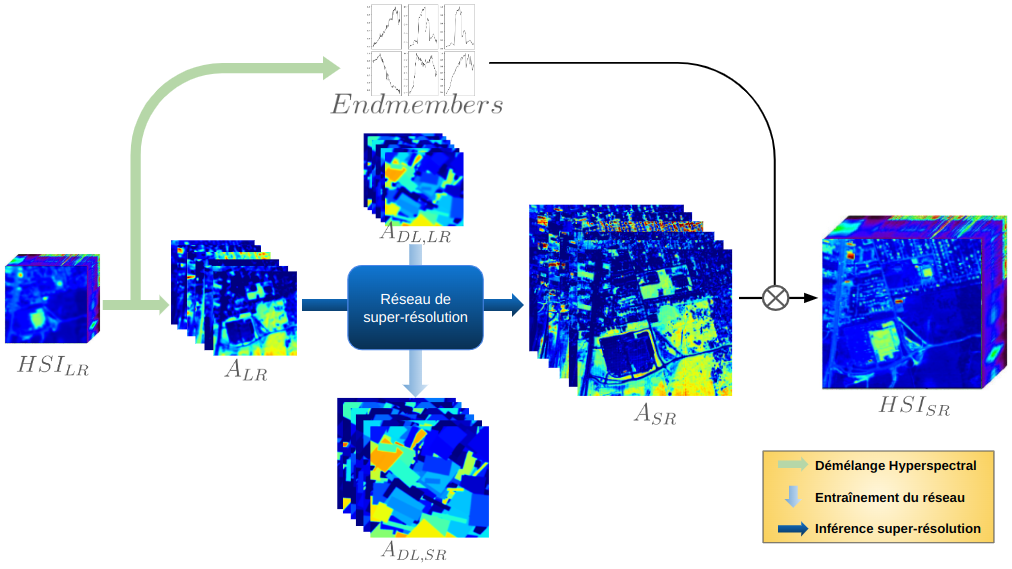}
    \caption{Structure de la méthode proposée: Le réseau de super-résolution est entraîné avec des paires synthétiques $(A_{DL,SR},A_{DL,LR})$ puis est utiliser pour super-résoudre $A_{LR}$ en $A_{SR}$}
    \label{fig:Structure}
\end{figure}

À partir de cette décomposition, l'idée principale de notre méthode consiste à générer un grand nombre de cartes d'abondances synthétiques $A_{DL,LR} \in \mathbb{R}^{N \times h \times w}$ et $A_{DL,HR} \in \mathbb{R}^{N \times H \times W}$ à partir de $A_{LR}$, en utilisant le modèle feuilles mortes. Les abondances haute résolution $A_{DL,HR}$ sont d'abord générées à l'aide de ce modèle, tandis que les abondances basse résolution correspondantes $A_{DL,LR}$ sont obtenues par l'application d'une réponse impulsionnelle (PSF) simulée par un flou gaussien et un sous-échantillonnage bicubique. Les paires $(A_{DL,HR}, A_{DL,LR})$ d'abondances synthétiques sont ensuite utilisées pour entraîner un réseau neuronal de super-résolution. Lors de la phase d'inférence, les abondances $A_{LR}$ de la HSI à super-résoudre sont fournies en entrée pour obtenir une estimation des abondances super-résolues $A_{SR}$. L'image hyperspectrale haute résolution finale $HSI_{SR}$ est obtenue en reconstruisant $HSI_{SR}(l,i,j) = \sum_{n=1}^{N} S(l,n) \cdot A_{SR}(n,i,j)$. La structure de notre méthode est résumée dans la Fig.~\ref{fig:Structure}.

Les sections suivantes décrivent en détail les étapes de la génération des cartes d'abondances synthétiques et du réseau de super-résolution.

\subsection{Images synthétiques en utilisant le modèle feuilles mortes}
\label{subsec:Synthetic Images using Dead Leaves model}

Le modèle feuilles mortes a été introduit pour la première fois en 1968 par Matheron pour modéliser les milieux poreux \cite{matheron1968modele, bordenave2006dead}, puis proposé plus tard comme un modèle pour les images naturelles~\cite{alvarez1999size, lee2001occlusion, gousseau2007modeling}. L'idée principale est de générer une image en superposant séquentiellement des formes aléatoires à des positions aléatoires, imitant ainsi le processus des feuilles mortes tombant d'un arbre. Les formes peuvent être définies à l'aide de n'importe quel modèle aléatoire, et leurs positions sont données par un processus ponctuel de Poisson stationnaire (des points répartis uniformément sur le plan). Le processus est itéré jusqu'à ce qu'un état stationnaire soit atteint, ce qui, en pratique, peut être obtenu à l'aide de techniques de simulation parfaite~\cite{kendall1999perfect} : chaque nouvelle forme est placée sous les formes précédentes, jusqu'à ce que l'image soit entièrement recouverte. Achddou et al. ont récemment montré qu'il était possible de super-résoudre des images naturelles avec un réseau entraîné uniquement à partir d'images synthétiques de feuilles mortes \cite{achddou2023fully}.

\begin{figure}
    \centering
    \includegraphics[width=1\linewidth]{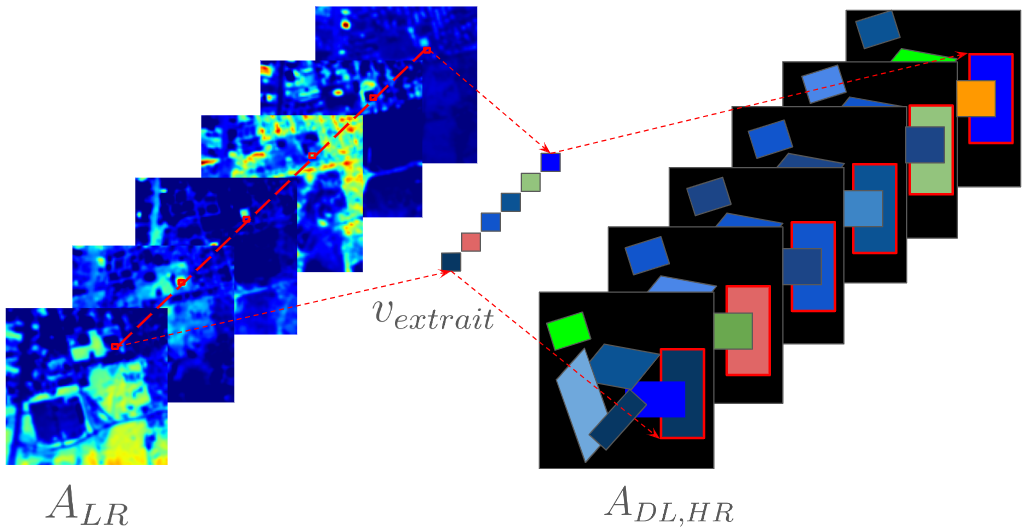}
    \caption{Une illustration de génération d'abondances synthétiques à la sélection de la $i^{ème}$ colonne de valeur $v_{extrait}$ pour les $A_{DL,HR}$.}
    \label{fig_DL_exemple}
\end{figure}

\begin{figure}
    \centering
    \includegraphics[width=1\linewidth]{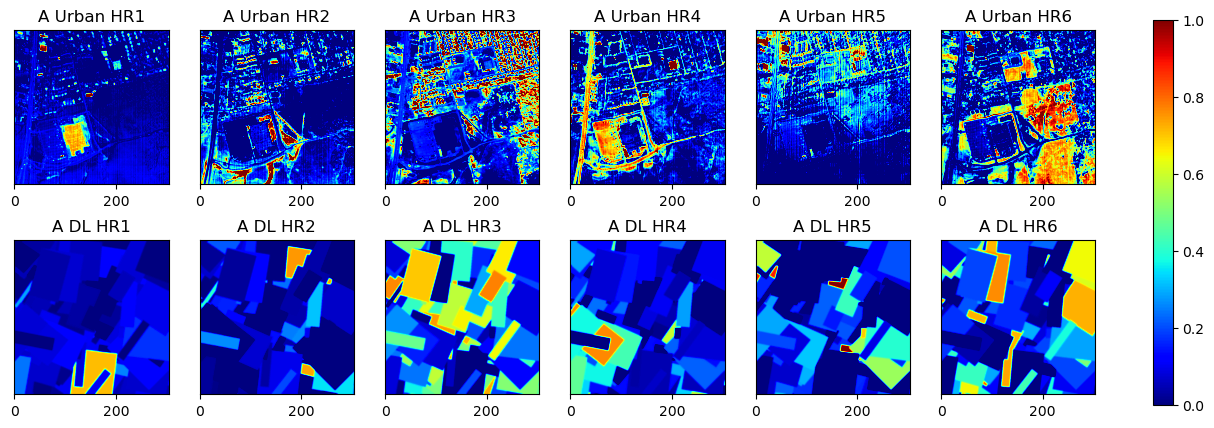}
    \caption{Comparaison entre les cartes d'abondances réelles d'urban (Haut) et les cartes d'abondances synthétiques de feuilles mortes (Bas).}
    \label{fig:compare_DL}
\end{figure}

\begin{figure*}[!htb]
    \centering
    \includegraphics[width=1\linewidth]{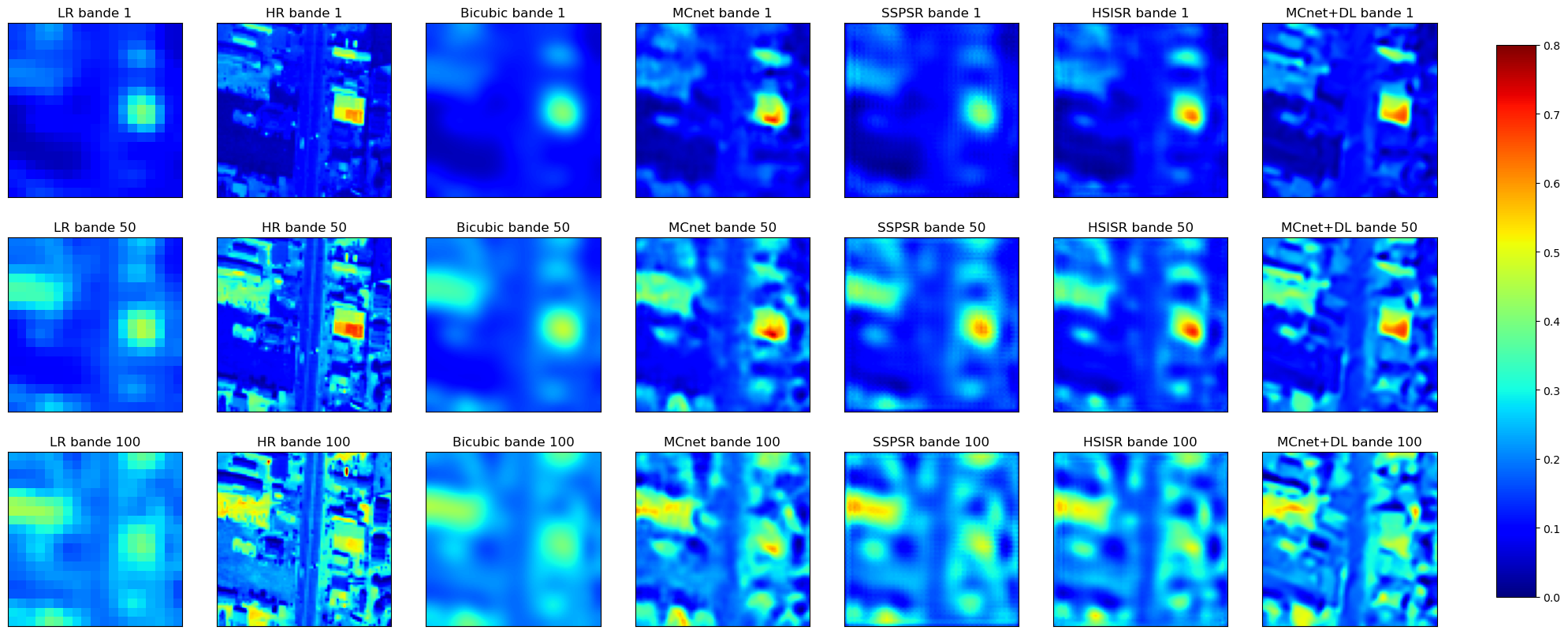}
    \caption{Comparaison visuelle entre la vérité terrain, LR, Bicubic, MCNet, SSPSR, HSISR et MCNet-DL sur un patch d'Urban à la bande n°1, 50 et 100}
    \label{fig:benchmark_fig}
\end{figure*}

Dans notre méthode, nous générons des cartes d'abondances synthétiques basées sur le modèle des feuilles mortes, en suivant le principe décrit précédemment. Nous considérons le contexte des HSIs urbains en télédétection, pour lequel l’utilisation de feuilles rectangulaires semble naturelle. Plus précisément, chaque feuille est définie par un vecteur $(a, b, \theta, V, x, y, z)$, où $a$ et $b$ représentent les dimensions d'une feuille rectangulaire, $\theta$ l’angle de rotation, $V \in [0, 1]$ la valeur associée à la feuille, $x$ et $y$ sont les coordonnées spatiales du centre de la feuille sur les abondances synthétiques et $z \in \{1, \ldots, N\}$ sa position dans la dimension des matériaux.

Afin de respecter les statistiques réelles des abondances dans le cadre du modèle des feuilles mortes, nous avons choisi de conserver, pour chaque dépôt de feuille, une même forme $(a, b, \theta)$ et une même position spatiale $(x,y)$ sur l’ensemble des dimensions des matériaux $z = \{1, \ldots, N\}$. Les valeurs des feuilles $V$ pour chaque carte d’abondance synthétique sont obtenues en sélectionnant aléatoirement un pixel contenant une répartition des matériaux $v_{extrait}$ directement à partir de $A_{LR}$ (voir illustration à la Fig. \ref{fig_DL_exemple}). Cette approche permet d'obtenir des proportions réalistes de matériaux dans les cartes synthétiques, comme illustré dans la Fig. \ref{fig:compare_DL}.

\subsection{Le réseau de super-résolution: MCNet}
\label{subsec : MCNet}

Le réseau de super-résolution MCNet (pour \textit{Mixed 3D/2D Convolutional Network}) est un CNN supervisé de l'état de l'art, initialement conçu pour super-résoudre directement des images hyperspectrales \cite{li2020mixed}. Comme son nom l'indique, il est composé de couches de convolution à la fois en 2D et en 3D. Les couches de convolution 2D sont conçues pour extraire des caractéristiques spatiales, tandis que celles en 3D permettent d'extraire simultanément des informations spatiales et spectrales. Afin de l'intégrer à notre architecture, nous fixons le nombre de canaux à $N$, correspondant au nombre de matériaux présents dans la scène.

\section{Résultats expérimentaux}
\label{sec: Results}

\subsection{Jeux de données \& Setup}
\label{subsec: Dataset}

Le dataset Urban considéré est un jeu de données capturé par le capteur \textit{Hyperspectral Digital Image Collection Experiment} (HYDICE). Il contient $307 \times 307$ pixels pour $210$ bandes dans la gamme de $400 - 2500$ nm. Après avoir retiré les bandes bruitées et corrompues, $162$ bandes sont utilisables. Pour évaluer les performances de notre algorithme, nous travaillons avec un facteur de super-résolution de $4$. La $HSI_{LR}$ est obtenue à l'aide d'une PSF simulée, en appliquant un flou gaussien suivi d'un sous-échantillonnage. L'écart-type du filtre gaussien est fixé à $\sigma = 4$, et sa taille est tronquée à $6 \sigma$ afin de respecter le critère de Shannon-Nyquist. Pour la phase d'entraînement, nous générons 5 000 cartes d'abondances $A_{DL,HR}$ synthétiques de taille $6 \times 307 \times 307$ selon le protocole décrit ci-dessus, et les cartes $A_{DL,LR}$ sont obtenues avec la même PSF.

Le réseau est entraîné pendant 200 époques en utilisant la fonction de perte L1 et l'optimiseur Adam, avec un taux d'apprentissage fixe de 0.0001. Nous évaluons les résultats \cite{aburaed2023review} avec le Peak Signal-to-Noise Ratio (PSNR), le Spectral Angle Mapper (SAM) et l'Erreur Relative Globale Adimensionnelle de Synthèse (ERGAS).

\subsection{Comparaison avec l'état de l'art}
\label{subsec: comparison}

Nous comparons notre méthode 
avec d'autres méthodes SISR de l'état de l'art (SOTA): MCNet\cite{li2020mixed}, SSPSR\cite{jiang2020learning}, et HSISR\cite{li2022hyperspectral}. Nous ajoutons également l'interpolation bicubique à notre baseline. Il convient de noter que toutes ces méthodes SOTA sont supervisées. Pour obtenir les données d'entraînement, nous découpons Urban en 16 patchs de taille $162 \times 76 \times 76$. Pour chaque méthode SOTA, 16 entraînements indépendants sont réalisés, où, à chaque fois, l'un des 16 patchs est réservé à l'évaluation des performances, tandis que les 15 autres servent à l'entraînement. Il est important de noter qu'une telle procédure, reposant sur une découpe artificielle des données, permet de favoriser ces méthode SOTA en leur offrant des conditions d’apprentissage optimales. Cependant, cette configuration est impraticable en conditions réelles. Enfin, les résultats numériques obtenus pour chaque méthode et chaque métrique sur chaque patch sont moyennés afin de permettre la comparaison.

Notre méthode (dénommée MCNet-DL dans les résultats) est uniquement entraînée sur le dataset de 5 000 cartes d'abondances Feuilles Mortes. Bien que notre méthode non supervisée permette une évaluation sur l'ensemble de l'image, nous utilisons le même processus de découpe pour nous adapter aux méthodes SOTA. Le Tableau \ref{tab:benchmark_tab} montre les métriques moyennes pour les 16 patchs évalués pour chaque méthode, y compris la nôtre. La Fig.\ref{fig:benchmark_fig} montre les résultats visuels pour un patch. Quantitativement, les trois métriques s'accordent sur la performance de notre méthode par rapport aux méthodes SOTA. De plus, en surpassant MCNet appliqué sur les patches de la HSI, cela démontre la pertinence de l'approche par données synthétiques. Visuellement, la méthode proposée permet de retrouver beaucoup plus de détails dans la HSI par rapport à l'interpolation bicubique,  SSPSR et HSISR. Par rapport à la méthode MCNet, notre reconstruction apparaît plus précise.

\begin{table}
    \centering
    \caption{Moyenne des résultats quantitatifs pour Bicubic, MCNet, SSPSR, HSISR et MCNet-DL sur 16 patchs d'Urban}
    
    \resizebox{\columnwidth}{!}{
    \begin{tabular}{c|c|c|c|c|c}
         & Bicubic & MCNet  & SSPSR & HSISR & MCNet-DL\\
         \hline
         mPSNR $\uparrow$& 24.17 & 25.56 & 24.56 & 25.24 &\textbf{26.69} \\
         mSAM $\downarrow$& 19.35 & 16.52 &18.21 & 17.18 & \textbf{14.53} \\
         mERGAS$\downarrow$& 10.31 & 8.72 & 9.82 & 9.12 & \textbf{7.60} \\
    \end{tabular}
    }
    \label{tab:benchmark_tab}
\end{table}

\section{Conclusion}
\label{Conclusion}
Dans cet article, nous avons proposé une méthode non supervisée pour la super-résolution d’images hyperspectrales en télédétection. L’idée principale de notre approche consiste à entraîner un réseau de super-résolution supervisé uniquement à l’aide d’images synthétiques de type feuilles mortes, ce qui permet de pallier le manque de jeux de données, limitation couramment rencontrée dans les méthodes de super-résolution. Nous montrons la supériorité de l'approche par rapport à plusieurs  approches de l’état de l’art. Par rapport aux travaux que nous avions présentés dans \cite{xu2024unsupervised}, l’étape de démélange hyperspectral est désormais pleinement intégrée dans la procédure globale de super-résolution, ce qui conduit à une méthode de super-résolution complète, efficace et non supervisée.

\textbf{Remerciements:} Ce travail a été partiellement soutenu par l'Agence de l'Innovation de Défense – AID - via Centre Interdisciplinaire d’Etudes pour la Défense et la Sécurité – CIEDS - (project 2023 - ALIA)

\begingroup
\small 
\bibliography{Biblio}
\nocite{*}
\endgroup


\end{document}